# Fluorescence coupling to plasmonic nanoparticles

Gernot Schaffernak, Christian Gruber, Joachim R. Krenn,
Markus Krug, Marija Gašparić, Martin Belitsch, Andreas Hohenau
Institute of Physics, University of Graz, Universitätsplatz 5, 8010 Graz, Austria

## ABSTRACT

The combination of single photon emitters (quantum dots) and tailored metal nanoparticles with defined size and shape allows a detailed study of the interaction between light and matter. The enhanced optical near-field of the nanoparticles can strongly influence the absorption and emission of nearby fluorescent quantum dots. We show that a controlled spatial arrangement enables the analysis and understanding of polarization dependent coupling between a metal nanoparticle and few or single fluorescent quantum dots. Modifications in the fluorescence spectrum and lifetime are analyzed and compare well with simulations.

The reduction of the fluorescence lifetime in such systems is usually in the order of 3-10. However, much larger reductions are to be expected if the quantum dots are positioned in a nanometric gap between two plasmonic nanoparticles, eventually leading to hot luminescence. We approach this regime experimentally and present first results from lithogaphically fabricated gold particle-pairs with controlled gap widths in the range of 1-20nm.

## 1. INTRODUCTION

Both, fluorescence and plasmonic excitations are interesting and important topics in nanooptics. Their combination allows for an investigation of the energy transfer between photons and surface plasmons. Recently, the controlled positioning of quantum dots (QDs) allowed for a direct measurement of the coupling of dipolar QD emission to surface plasmons on a silver nanowire and vice versa[1,2]. The polarization direction dependence of the coupling to surface plasmons on the nanowire[2,3,4] was an important result for the understanding of surface plasmon modes. Changing the QD position relative to the nanowire and measuring the fluorescence decay time allows a direct mapping of the local density of states (LDOS), (figure 1).

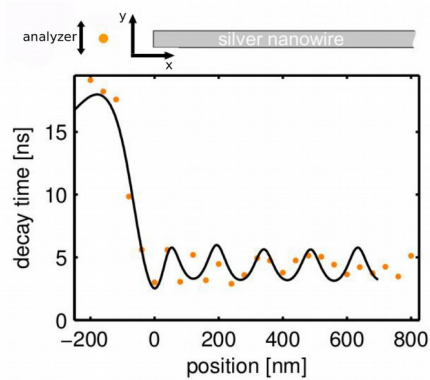

Figure 1. QD (600 nm emission) decay time measured at different x-positions along a 2 μm long silver nanowire. The analyzer polarization is oriented in y-direction. Fabrication was done with 2-step electron-beam lithography with a 10 nm $SiO_2$ layer between QD and nanowire (100 nm width, 50 nm height). Adapted from reference 4.

A further increase of the LDOS and thus the coupling strength between fluorophore and the plasmonic particle should be possible, if the emitter is positioned in a small gap between two metal nanoparticles. In the following, first attempts to lithographically fabricate reproducible and controlled gap widths between particle pairs are presented and analyzed for their local optical properties.

## 2. NANO-BURGER

There are several possibilities to create nanogaps of controlled size below 10 nm, including electron-beam lithography with high energies[5], sphere-on-plane setup[6] and chemically grown nanoparticles with molecule-layers as spacer material[7]. We found the combination of electron-beam lithography fabrication and layered evaporation of a gold nanoparticle, several nm of dielectric spacer medium and another gold particle interesting for two reasons. First, the strong plasmonic coupling between the particles leads to large shifts in the plasmon resonance frequencies and second, the electron beam evaporation of the dielectric allows for a defined spacer thickness even below 5 nm. Figure 2 shows a sketch of the layer structure and explains the charge distribution of two coupled disks at the resonance frequencies of the bonding and antibonding mode.

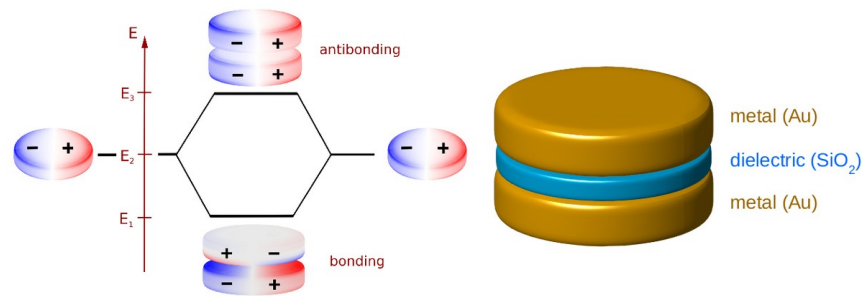

Figure 2. Left: Resonance modes of the nano burger. For the antibonding solution the dipolar modes in the two disks are in phase, while for the bonding solution they have opposite phase. Right: Concept of the nano-burger: gold, $SiO_2$ and gold layer.

### 2.1 Fabrication

Glass substrates with conductive ITO (indium tin oxide) coating are cleaned, spin coated with a PMMA (polymethylmethacrylat, Allresist AR-P 671.02) layer of about 120 nm thickness and baked on the hot plate for 5 minutes at 180 °C. Designed arrays of gold disks with 100 nm diameter and a grating constant of 300 nm were exposed with 10 kV accelerating voltage in the electron-beam lithography process. The voltage was chosen in order to ensure the presence of an undercut in the developed PMMA structures, leading to a clean lift-off afterwards. Development in IPA:MIBK = isopropanol:isopropylacetone (13:2) for 30 s was followed by 30 s stopping in IPA and 20 seconds plasma cleaning. The gold layers of 30 nm thickness were deposited by thermal evaporation, while the spacer was evaporated by electron-beam evaporation. Finally, the PMMA mask was removed in aceton.

### 2.2 Extinction spectra

In total, six samples are analyzed here. The first one includes arrays of gold nanodisks with 30 nm height and serves as a reference. The following four samples were prepared with the nano-burger architecture with spacer layers of 20 nm, 10 nm, 5 nm and 1 nm. The last sample consists of gold nanodisks with 60 nm height and mimics the case when the spacer layer does not separate the disks properly. Figure 3 shows the measured extinction spectra and in comparison the extinction spectra that were calculated with the MNPBEM[8] toolbox, embedding the disks in a homogeneous dielectric surrounding with an average refractive index 1.27. The simulation results show characteristic extincion spectra with two resonances corresponding to the bonding and antibonding mode, which can be concluded from the surface charge distribution at these resonance wavelengths. With decreasing gap between the disks the simulation predicts a slight blueshift for the antibonding mode and a distinct redshift for the bonding mode. The splitting into the bonding and antibonding mode was observed also in the experiments for the nano-burgers with 20 nm, 10 nm and 5 nm of spacer layer. As expected, the bonding mode shifts strongly to larger wavelengths, while no clear trend could be observed for the antibonding mode, as a result of small particle-size deviations between different samples. The lower the interparticle distance, the less the bonding mode can be excited and its extinction peak de facto disappears for only 1 nm of spacer layer.

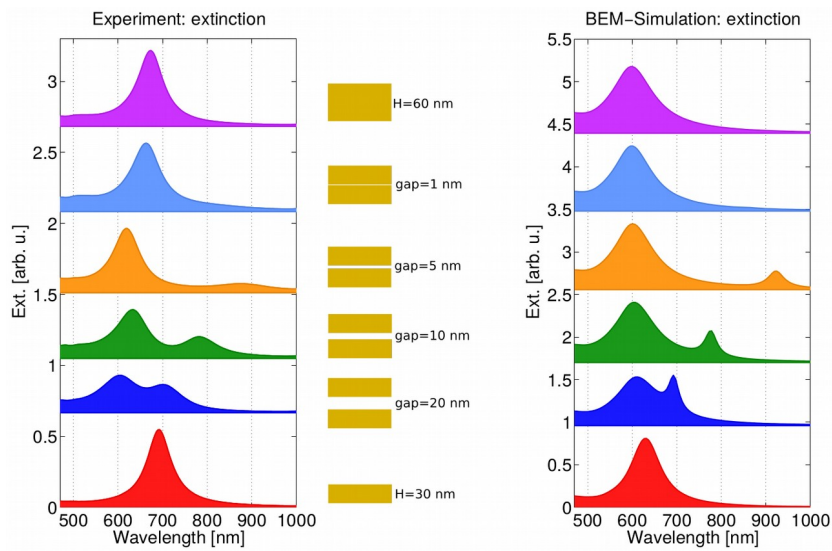

Figure 3. Optical extinction spectra of gold disk and nano-burger arrays. BEM-simulations of the same geometries.

**2.3 AFM tests on the spacer layers**

The sample with 5 nm of $SiO_2$ as spacer obviously has clearly separated disks as can be seen from the mode splitting in the spectra. For smaller dielectric layers the spectral information is not conclusive as in the visible spectrum no obvious features are expected from the simulations. On SEM images a layer of 1 nm $SiO_2$ can be distinguished from an underlying gold surface, but the image does not allow to judge about small electric defects (short cuts) in the layer (figure 4).

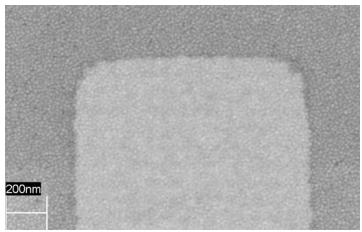

Figure 4. SEM image at 3 kV of a 1 nm thick $SiO_2$ stripe on a gold surface. The bright region corresponds to the SiO2.

We decided to check if the small spacer layers are good isolators with conductive atomic force microscopy (AFM). A potential was applied between the tip and the test substrate: a gold surface and lithographed disks with thin layers of $SiO_2$ and 30 nm of gold on top. While the freshly evaporated gold substrate alone showed very good conductivity, the contact resistance increased strongly for the sample with the disks after the liftoff process. We note that voltages of 5 to 8 V were needed to get a clear signal, compared to about 1 V for the flat gold film. The voltage signal was stored together with the height information in the AFM software. Results for 1 nm and 2 nm dielectric layers are shown in figure 5.

### 3. DIPOLE DECAY RATE

As a preparative work to successfully build a system with fluorophores in the spacer layer, simulations have been done to estimate the coupling strength for various particle geometries, spacer heights and emission wavelengths of the dipolar emitter. As an example, the total and radiative decay rate ratios are given for a 100 nm diameter nano-burger with 10 nm gap between the disks. The free-space decay rate is set to 1 such that the given ratios are equivalent to the partial LDOS for the given polarization direction. The extinction simulation revealed two resonance frequencies for this system around 600 nm and 780 nm and the calculation results for both are given in figure 6.

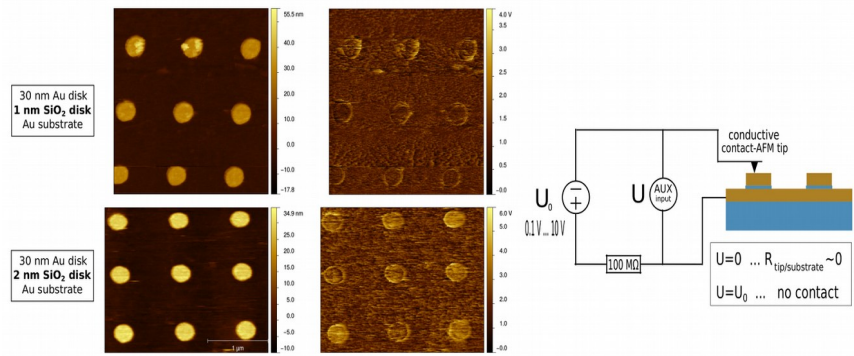

Figure 5. AFM images with height signal to the left and voltage signal from conductive AFM in the middle. Upper images testing the 1 nm layer, lower images the 2 nm layer, showing less electric contact over the disks. Setup sketch to the right.

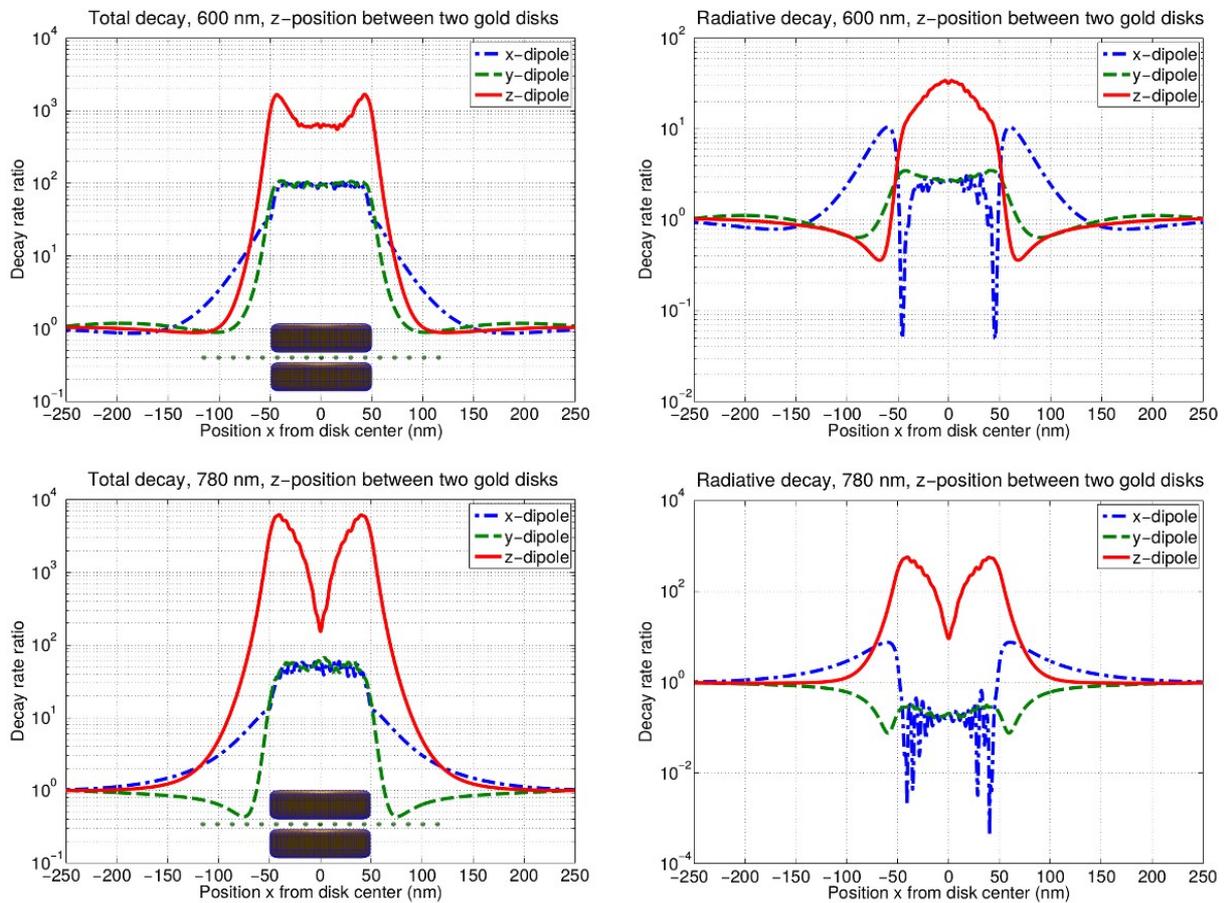

Figure 6. Total and radiative decay rate ratio for 3 orientations of the dipole in x, y and z direction. Inset sketching the model used for the boundary element method and the positions of the dipole emitter. 600 nm emitter on top, 780 nm below.

The large calculated decay rate increase of factor $10^3$ to $10^4$ is a promising result and including the fluorescence source into the dielectric spacer layer is actual work in progress. We are expecting even larger coupling strengths between the emitter and the nanoparticles than for the system of the QD near to a single nanoparticle. The z-dipole results suggest a good detectability with quantum yields between 5 and 10 % for emitters in the gap region.

We acknowledge funding from FWF project P 25034 and NAWI Graz.